\documentclass[conference]{IEEEtran}
\usepackage{amsmath,amssymb, amsthm}
\usepackage{graphicx}
\usepackage{verbatim}
\usepackage{subfig}
\usepackage{wrapfig}
\usepackage{setspace}
\usepackage{ifthen}
\usepackage{multicol}
\usepackage{cite}
\usepackage{enumerate}
\usepackage[usenames,dvipsnames]{color}
\usepackage[font=normalsize]{caption}

\usepackage{fullpage}
\DeclareMathOperator{\bone}{\mathbf{1}}

\def\ba{\begin{array}}
\def\ea{\end{array}}

\newcommand{\beq}{\begin{equation}}
\newcommand{\dd}{\partial}
\newcommand{\eeq}{\end{equation}}
\newcommand{\bq}{\begin{eqnarray}}
\newcommand{\eq}{\end{eqnarray}}
\newcommand{\bqn}{\begin{eqnarray*}}
\newcommand{\eqn}{\end{eqnarray*}}
\newcommand{\bee}{\begin{enumerate}}
\newcommand{\eee}{\end{enumerate}}
\newcommand{\bi}{\begin{itemize}}
\newcommand{\ei}{\end{itemize}}

\newtheorem{theorem}{Theorem}
\newtheorem*{theoremNoNum}{Theorem}

\newtheorem{lemma}{Lemma}

\newboolean{showcomments}
\setboolean{showcomments}{true}

\newcommand{\bose}[1]{  \ifthenelse{\boolean{showcomments}}
{ \textcolor{Red}{(Bose says:  #1)}} {}  }
\newcommand{\desmond}[1]{\ifthenelse{\boolean{showcomments}}
{ \textcolor{Red}{(Desmond says:  #1)}}{}}
\newcommand{\steven}[1]{\ifthenelse{\boolean{show-comments}}
{ \textcolor{Red}{(Steven says: #1)} } {} }
\newcommand{\adam}[1]{\ifthenelse{\boolean{showcomments}}
{ \textcolor{Red}{(Adam says: #1)} } {} }

\IEEEoverridecommandlockouts
\title{\vspace{.20in}{The Role of a Market Maker \\ in Networked Cournot Competition}}
\author{Subhonmesh Bose, Desmond Cai, Steven Low, Adam Wierman\\ California Institute of Technology
\thanks{Emails: \{boses@, wccai@, slow@cms., adamw@cs.\}caltech.edu. This work was supported in part by NSF grants CNS 0911041, ECCS 1253516, CNS-1319820, EPAS-1307794 and ARPA-E grant DE-AR0000226, Southern California Edison, Nat. Sc. Council of Taiwan, R.O.C. grant NSC 101-3113-P-008-001,
Resnick Inst., and Okawa Found.}
}
\begin{document}

\maketitle

\begin{abstract}
We study the role of a market maker (or market operator) in a transmission constrained electricity market. We model the market as a one-shot networked Cournot competition where generators supply quantity bids and load serving entities provide downward sloping inverse demand functions. This mimics the operation of a spot market in a deregulated market structure. In this paper, we focus on possible mechanisms employed by the market maker to balance demand and supply. In particular, we consider three candidate objective functions that the market maker optimizes -- social welfare, residual social welfare, and consumer surplus. We characterize the existence of Generalized Nash Equilibrium (GNE) in this setting and demonstrate that market outcomes \emph{at equilibrium} can be very different under the candidate objective functions.
%In this paper, we study the role of the market maker (or market operator) in an electricity market subject to network capacity constraints. We model the market as a networked Cournot model and we show that equilibrium always exists if network capacity constraints are not binding, but might not exist if network capacity constraints are tight. We analyze three market clearing rules - social welfare maximization, residual social welfare maximization, and consumer surplus maximization. and we show that different market clearing rules can lead to significantly different market equilibria behavior.
\end{abstract}


%\onehalfspacing
%%%%%%%%%%%%%%%%%%%%%%%%%%%%%%%%%%%%%%%%

\section{Introduction}

In the 1990s, the Federal Energy Regulatory Commission (FERC) began to deregulate electricity markets in various states by replacing cost-of-service regulated rates with market-based prices. The goal was to increase competition among generators and lower prices to end-consumers. However, the consequences of deregulation were unexpected; in 2000 and 2001, market manipulations led to the California electricity crisis which involved multiple large-scale blackouts and skyrocketing prices~\cite{sweeney2008california}. It is estimated that about \$5.55 billion was paid in excess of costs between 1998 and 2001 alone~\cite{borenstein2002measuring}. Subsequently, various measures were introduced in the markets to curb such behavior. Nevertheless, market manipulation continues to exist. For instance, JP Morgan was fined \$410 million for market manipulation in electricity markets in California and the Midwest from September 2010 to November 2012~\cite{jpmorgan}. Clearly, market power remains a major issue in electricity markets today.

One reason for this is that electricity markets are complex and operate on multiple time-scales. Power delivered at a particular instant of time is first procured months (or even years) ahead via long-term bilateral contracts between generators and retailers. Between one week and one day ahead of delivery, generators and retailers begin to trade in centralized electricity markets to clear imbalances. These centralized markets typically operate over multiple stages to allow market participants to exploit the increased information about supply and demand closer to delivery. The procedures for each stage are similar -- generators and retailers submit offers and bids respectively and the market operator clears the market by solving a centralized dispatch problem to minimize system costs subject to operating constraints. Payments are calculated based on locational marginal prices (LMP) which are designed to reflect local costs of generation. Clearly, the \emph{market clearing mechanism} employed by the market operator has a significant impact on the incentives and bidding behavior of market participants.

\emph{Our goal in this paper is to study the role of the market maker (or market operator) on the efficiency of electricity markets.} In particular, we study the impact of the market clearing rule on the strategic incentives of generators. We formulate a competitive model between multiple generators and a single market maker on a power network. We model the bidding behavior of the generators using a networked Cournot model and we assume that the market maker clears the market to optimize a specific regulatory objective subject to the DC power flow equations and line constraints. In particular, we consider three natural market maker objectives -- social welfare, residual social welfare, and consumer surplus. The social welfare is the consumers' utility less the costs of generation; the residual social welfare is the consumers' utility less the revenue of generators; and the consumer surplus is the consumers' utility less their payment. Note that residual social welfare is not necessarily equal to the consumer surplus since electricity markets are not necessarily budget balanced. The latter is a byproduct of nodal pricing -- generators receive local prices which might be different from the price at the remote location where their power is delivered.

\subsection{Contributions of this paper}

We make two main contributions: (i) we characterize the existence of equilibria under each of the three market maker objectives, and (ii) we show that, when equilibria exists, the equilibrium flow could be completely different under the three objectives. Our results highlight the importance of designing the market in a way that takes into account strategic generator behavior and physical system constraints. The equilibrium concept we adopt in this paper is known as Generalized Nash Equilibrium (GNE). As will be clear in Section~\ref{sec:formulation}, the strategy set of the market maker depends on the actions of generators, and so the conventional Nash equilibrium framework does not apply to our setting. Hence, we resort to GNE which is an extension of Nash equilibrium for such settings.

Our first main result is that a GNE always exists under the social welfare and residual social welfare objectives but it might not exist under the consumer surplus objective. For the latter, we provide a simple 2-node example under which GNE does not exist. Our proof shows that one of the key factors that leads to non-existence of GNE is that the consumer surplus is not a concave function of the market maker's decision variable. Non-existence of equilibria could have numerous negative implications on market efficiency, e.g. more volatile prices leading to higher risk premium that eventually translates into higher costs for consumers. Also, market power measures might need to be adjusted to use longer-term metrics in order to account for the unreliable observations of market outcomes (e.g. see~\cite{bose2014}).

Our second main result shows that, when equilibria exists, the market outcomes could differ significantly under the three regulatory objectives. In particular, we focus on a 2-node example and show that the equilibrium flow could be positive with social welfare maximization, zero with residual social welfare maximization, and negative with consumer surplus maximization. Hence, although all three regulatory objectives attempt to maximize consumer benefit, the exact methodology by which system costs are reflected in the objective impacts how generators behave in the market and determines the resulting equilibrium and system efficiency.

\subsection{Related literature}

Electricity markets are challenging to model and analyze due to the multiple time-scales, non-convex generation costs, network constraints and generation supply constraints. Nevertheless, there is a sizable literature focused on analyzing the key strategic incentives of generators. The models that have been used can be largely classified into two categories -- supply function competition and Cournot competition. In both approaches, it is common to assume that demand is exogenous and focus on analyzing the resulting strategic game among generators. Here, we briefly review prior work using supply function and Cournot competition in single-stage settings. We recognize that there is also significant work in multi-stage models, but we do not discuss that here as forward contracting is not the focus of the current paper.

\emph{Supply function competition:} Introduced by Klemperer et al. in~\cite{klemperer1989supply}, the key feature of supply function competition is that firms (or generators) compete by choosing supply functions specifying how much power it is willing to supply at each price. This model is appealing due to its similarity to how electricity markets operate in practice where generators typically submit step-wise increasing offer functions. Hence, this model has been frequently used both analytically and numerically to obtain insights on generator behavior~\cite{green1992competition, green1996increasing, bunn2001agent, bunn2003evaluating, baldick2001capacity,rudkevich2005supply,baldick2004theory}. In certain cases, strong theoretical results were obtained by restricting the functional form of the supply functions to a parameterized class~\cite{green1996increasing,rudkevich2005supply,baldick2004theory,johari2011parameterized} (typically affine or logarithmic). More recent work has analyzed supply function competition in settings with network transmission constraints~\cite{xu2002transmission, xu2007transmission, wilson2008supply}. However, to our knowledge, no work has addressed the role of the market maker under supply function competition.

\emph{Cournot competition:} Cournot competition is a well-known competitive model in economics dating back to 1838~\cite{varian1992microeconomic, mas1995microeconomic}. In contrast with the supply function approach where generators submit an offer function, in Cournot competition, generators submit a single quantity specifying how much they are willing to supply at any price. Hence, this formulation amounts to generators specifying a supply function with zero price elasticity. Although this offer model is significantly different from how electricity markets operate in practice, it was found that the Cournot model often provides good explanations of observed price variations~\cite{ventosa2005electricity, willems2009cournot}. Further, the Cournot model is appealing due to its tractability, e.g. bounds on the loss in system efficiency due to strategic behavior have been obtained~\cite{johari2005efficiency, tsitsiklis2012efficiency, tsitsiklis2013profit}.

\emph{Networked Cournot competition:} Cournot competition has also been applied to settings with network transmission constraints~\cite{neuhoff2005network, barquin08cournot, barquin2005cournot, yao2004computing, yao2007two, yao2008modeling, jing1999spatial}. Such frameworks have also been applied to domains outside electricity within a broader framework referred to as networked Cournot competition~\cite{ilkilic2009cournot}. However, the results in~\cite{ilkilic2009cournot} are not directly applicable to electricity markets because they ignore network flow constraints. To our knowledge, in both non-networked and networked Cournot competition, no work has studied the role of the market maker which is the main focus of this paper.

\section{Problem Formulation}
\label{sec:formulation}

Our goal in this paper is to understand how the decision of the market maker impacts the strategic incentives and the resulting market equilibrium of generators in an electricity market. Hence, we model the market as a game between two entities: generators located at different nodes of the network, and a market maker that balances demand and supply. Since nodal pricing is a key feature in many electricity markets, we seek to capture this feature in our model by having generators and demand face different prices depending on their location in the network.

\subsection{Notation}

Let $\mathbb{R}$ denote the set of real numbers and $\mathbb{R}_+$ denote the set of non-negative real numbers. For any two vectors $u$, $v$ of the same size, we say $u \geq v$ if the vector $u - v$ is element-wise non-negative. Also, let $\bone$ denote the vector of all ones of appropriate size. For any vector $v \in \mathbb{R}^n$, we denote its transpose by $v^{\top}$. We also let $v_{-i} = (v_1,\ldots,v_{i-1},v_{i+1},\ldots,v_n)$ denote the vector of all elements other than the $i$-th element.

\subsection{Network model}

We consider a power network with $n$ nodes $1, 2, \ldots, n$ and $\ell$ edges. Each node $k$ has a generator $G_k$ that supplies a quantity of power $q_k \geq 0$ and incurs a cost $c_k q_k^2$ for some $c_k > 0$. We assume that demand at each node $k$ can be represented by a linear demand function:
$$p_k(d_k) := a_k - b_k d_k,$$
for some $a_k > 0$ and $b_k \geq 0$. Here, $p_k(d_k)$ is the price that demand at node $k$ is willing to pay as a function of the quantity of power $d_k$ it receives. This form of demand function is a common assumption in economics~\cite{varian1992microeconomic} and prior studies of electricity markets models~\cite{yao2004computing,yao2007two,yao2008modeling} and corresponds to an aggregate consumer having a quadratic utility function. We also assume that all demand functions are fixed and known to all market participants, which is reasonable when demand is highly predictable.

We assume that there is a single market maker $M$ that balances supply and demand by choosing \emph{re-balancing quantities} $r_k$ at each node such that demand at node $k$ receives a quantity:
$$ d_k := q_k + r_k.$$
At each node $k$, the market maker charges the demand and pays the generator at a price $p_k ( q_k + r_k )$. This model for nodal pricing is motivated by prior studies of electricity markets, e.g.~\cite{yao2004computing, yao2007two, yao2008modeling}.

%%%%%%%%%%%
\begin{figure*}[!t]
\begin{align}
W_{soc}(q, r)
& \ := \ \sum_{1\leq k \leq n} \left( \int_{0}^{q_k + r_k} p_k(w_k) d w_k  \ - \ \tilde{c}_k(q_k) \right).
\label{eq:soc.G} \\
W_{res} (q, r)
& \ := \ W_{soc}(q, r) - \sum_{1\leq k\leq n} \pi^G_k(q, r) \ \  = \ \sum_{1\leq k \leq n}  \Bigg(  \int_{0}^{q_k + r_k} p_k(w_k) d w_k  - q_k p_k(q_k + r_k) \Bigg). \label{eq:res.G} \\
W_{con} (r, q)
& \ := \ \sum_{1\leq k \leq n}  \Bigg(  \int_{0}^{q_k + r_k} p_k(w_k) d w_k  - (q_k + r_k ) p_k(q_k + r_k) \Bigg). \label{eq:con.G}
\end{align}
\hrulefill
\vspace*{2pt}
\end{figure*}
%%%%%%%%%%%%

Let the vector $q := (q_1, q_2, \ldots, q_n)$ denote the production quantities of the generators and the vector $r := (r_1, r_2, \ldots, r_n)$ denote the re-balancing quantities chosen by the market maker. We assume that the market maker chooses the vector $r$ of re-balancing quantities subject to the following constraints:
\begin{enumerate}[(i)]
\item Demand at each node is non-negative, i.e., $q + r \geq 0$.
\item Power flow on each transmission line respects the line limits, i.e., $-f \leq -H r \leq f$, where $H \in \mathbb{R}^{\ell \times n}$ is the shift-factor matrix that relates the flows on all $\ell$ lines as a function of the power injection vector $-r$ and $f \in \mathbb{R}^{\ell}$ is the vector of all line capacities.
\item Re-balancing quantities sum to zero, i.e., $\bone^{\top} r = 0$.
\end{enumerate}

Note that the set of allowable re-balancing quantities depends on the production quantities $q$. We denote the set of allowable re-balancing quantities by:
$$S^M (q) := \Bigg\{ r \in \mathbb{R}^n \ \left| \ q + r \geq 0, | H r | \leq f, \bone^{\top} r = 0 \right.\Bigg\}.$$
Figure~\ref{fig:2Node} shows an example of a 2-node network, which we study in detail in Section \ref{sec:regObj}.

We remark that the shift-factor matrix depends on the admittances of the transmission lines of the power network and encapsulates Kirchoff's laws \cite{stoft2002power}. This representation assumes a linearized DC power-flow model \cite{Purchala2005} for the network. Though widely used in the literature, this representation of the power flow equations has its limitations for power system operation, e.g., see \cite{stott2009dc}. However, in electricity markets, locational marginal prices are typically calculated using the DC power-flow model~\cite{li07LMP,caiso,nyiso}. Hence, this is a reasonable model for the purpose of studying generator bidding behavior in the market.

\subsection{Generator profit}

Within the context described above, the profit of generator $G_k$ is given by:
\begin{align}
\label{eq:profGen}
\pi_k^G (q_k, q_{-k}, r) := q_k p_k (q_k + r_k) - c_k q_k^2.
\end{align}
We assume that each generator seeks to maximize its profit~$\pi_k^G (q_k, q_{-k}, r)$ over its production quantity $q_k\in S^G_k$ where~$S^G_k = \mathbb{R}_+$ denotes the set of allowable production quantities of generator~$k$. That is, we assume that generators have infinite capacities.

This is a common assumption in prior studies of market power~\cite{green1992competition,green1996increasing}. The analysis of the case of finite generation capacities is clearly important, but it is left for future work.

Notice that, without the strategic market-maker and geographically distributed generators, this model reduces to the standard Cournot oligopoly in the microeconomics literature~\cite{varian1992microeconomic, mas1995microeconomic}.

\subsection{Market maker objectives}

Our focus in this paper is on the role of the market maker.   In electricity markets, the market makers are often regulatory authorities, e.g., ISOs; thus our goal is to study the role of market design in this regulatory framework.

The market maker designs we consider assume that the market maker maximizes some objective function $\pi^M(q, r)$ over the re-balancing quantities $r \in S^M(q)$. Note that the market maker is a regulatory authority and is free to choose a suitable payoff function. This is the \emph{market design} question of interest, and in this paper we analyze different candidates for the payoff function $\pi^M(q, r)$.

Specifically, inspired from the microeconomics literature~\cite{varian1992microeconomic, mas1995microeconomic}, we consider the following candidates for $\pi^M(q, r)$:
\begin{enumerate}[(a)]
\item \emph{Social welfare:} This is the net benefit to society. It refers to the consumers' utility less generation costs (also referred to as overall network utility). We denote it by $W_{soc}(q,r)$ in \eqref{eq:soc.G}. If generators are not strategic, this corresponds to the original optimal power flow formulation in \cite{carpentier1962contribution}.

\item \emph{Residual social welfare:} In practice, generator costs are unlikely to be known to the market maker. Hence, an alternative regulatory objective is to maximize the social welfare, less the profits of the generators. This is equivalent to the consumers' utility less the revenue of the generators. We denote it by $W_{res}(q,r)$ in \eqref{eq:res.G}.

\item \emph{Consumer surplus:} This is the net benefit to consumers. It refers to the consumers' utility less their payments. We denote it by $W_{con}(q, r)$ in~\eqref{eq:con.G}.
\end{enumerate}

We remark that at each node $k$, the amount paid by the consumers is $(q_k + r_k ) p_k(q_k + r_k)$, and the amount paid to the generator $G_k$ is $q_k p_k(q_k + r_k)$. Hence, the market is not necessarily budget-balanced. The difference between the total payment by demand and the total revenue of the generators has previously been referred to as \emph{merchandising surplus}~\cite{wu1996folk}. A consequence of the market not being always budget-balanced is that the residual social welfare is not necessarily equal to the consumer surplus. Hence, it is important to explore the impact of both objectives on the market.

\subsection{Competitive model}

Given the models of the generators and the market maker, we now need to model their interaction.  To do this, we consider a game with: (a) players $( G_1, G_2, \ldots, G_n, M)$; (b) strategy sets $(S^G_1, S^G_2, \ldots, S^G_n, S^M)$; and (c) payoffs $(\pi^G_1, \pi^G_2, \ldots, \pi^G_n, \pi^M)$, where $\pi^M$ is chosen to be one of the functions in $\{ W_{soc}, W_{res}, W_{con} \}$. Throughout, we assume that the game is feasible, i.e., the set $\{ (q \in \mathbb{R}^n_+, r \in \mathbb{R}^n) \ \vert \ (q, r) \in (S^G_k, 1\leq k \leq n, \ S^M(q)) \}$ is non-empty.

Since the strategy set $S^M(q)$ of the market-maker depends on the actions $q$ of the generators, we focus on a type of equilibrium known as \emph{Generalized Nash Equilibria} (GNE). Formally, an action profile $(q^*, r^*)$ constitutes a GNE if, for each $1 \leq k \leq n$, we have:
\begin{align*}
\pi^G_k (q_k^* , q_{-k}^*, r^* ) & \ \geq \ \pi^G_k (q_k , q_{-k}^*, r^* ) \text{ for all } q_k \in S^G_k, \\
\pi^M (q^*, r^* ) & \ \geq \ \pi^M (q^*, r ) \text{ for all } r \in S^M(q^*).
\end{align*}
This equilibrium concept was first introduced in 1952 by Debreu~\cite{debreu1952social}. It is an extension of Nash equilibrium where the strategy sets of players do not depend on the actions of the other players. We refer the reader to~\cite{facchinei2010generalized} for a detailed survey.

%Using Assumption \ref{ass:simple}, we have
%\begin{align}
% W_{soc}(q,r) &= \sum_{1 \leq k \leq n} a_k (q_k + r_k) - \frac{b_k}{2}(q_k + r_k)^2 - c_k q_k^2, \label{eq:soc}\\
% W_{res}(q,r) &= \sum_{1 \leq k \leq n} a_k r_k -  \frac{b_k}{2}(r_k^2 - q_k^2), \label{eq:res}\\
% W_{con}(q,r) &= \sum_{1 \leq k \leq n} \frac{b_k}{2}(q_k + r_k)^2 \label{eq:con}.
%\end{align}

%%%%%%%%%%%%%%%%%
\begin{figure}[!t]
\vspace{-20pt}
\centering
{\includegraphics[width=0.35\textwidth]{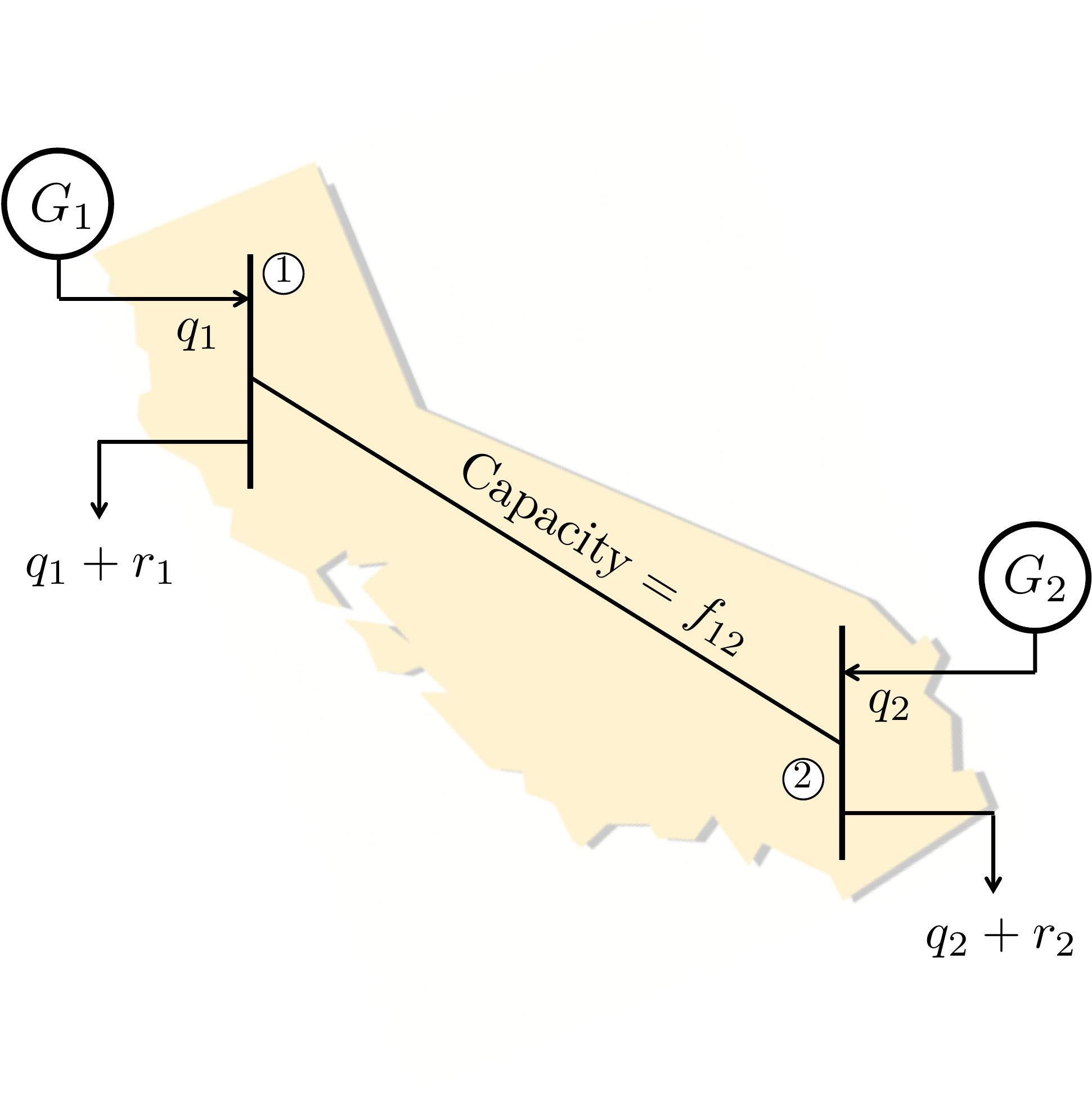}} \vspace{-20pt}
\caption{\emph{Example of a 2-node network. This example illustrates how the model in this paper can be used to study a caricature of the market in California. Here, northern and southern California are represented as two aggregate nodes connected by a transmission line - Path 15 - that is often congested~\cite{Chris}.}} \label{fig:2Node}
\end{figure}
%%%%%%%%%%%%%%%

\section{Existence of equilibrium}
\label{sec:exist}

Within the context of the model described in the previous question, we seek to investigate the following two questions in this paper:
\begin{itemize}
\item[1.] Does a GNE always exist for every each of the market maker objectives we have described, i.e., $\pi^M \in \{ W_{soc}, W_{res}, W_{con} \}$?
\item[2.] In the cases where a GNE exists, how do the market outcomes (in terms of flows, profits of generators and social welfare) differ for different market maker objectives?
\end{itemize}

We focus on the first question in this section and treat the second question in Section \ref{sec:regObj}.

The following is our main result on the existence of GNE.
\begin{theorem}
\label{thm:exist}
A GNE exists if $\pi^M = W_{soc}$ or $\pi^M = W_{res}$. However, a GNE may not exist if $\pi^M = W_{con}$.
\end{theorem}

The theorem shows that the market maker objective has a significant impact on the existence of a GNE in the market. One of the key factors that lead to non-existence of GNE is that the consumer surplus $W_{con}$ is not a concave function of the re-balancing quantities $r$. Hence, when $\pi^M = W_{res}$, the optimal re-balancing quantities are at the boundaries of the feasible set $S^M(q)$. When the generator production $q$ changes, the optimal re-balancing quantities $r^*$ could jump from one boundary point to another, i.e. it is not always continuous in $q$, especially when network capacity constraints are binding. Hence, there does not necessarily exist a fixed-point in $(q, r)$. In the proof, we explicitly construct an example that exhibits this phenomena using the 2-node network in \figurename~\ref{fig:2Node}.

Given Theorem \ref{thm:exist}, let us briefly emphasize the importance of choosing a regulatory objective that leads to existence of equilibria. Non-existence of equilibria could have numerous negative implications. It could lead to volatile market prices as the market oscillates between different outcomes which would increase the risk premium and the cost of forward contracting. Market power measures might need to be adjusted to use longer-term metrics in order to account for the unreliable observations of market outcomes (e.g. see~\cite{bose2014}). Further, more sophisticated models and equilibria concepts (e.g. repeated game models, dynamic equilibria) might have to be used in theoretical and empirical analysis of market behavior.

To prove the existence results in Theorem~\ref{thm:exist}, we use a result that can be traced back to Debreu~\cite{debreu1952social}. However, the version we apply is a slightly simplified statement given in~\cite{facchinei2010generalized, ichiishi1983game}. Below, we state Debreu's theorem before giving a proof of Theorem \ref{thm:exist}.

\begin{theoremNoNum}(Debreu \cite{debreu1952social})
Consider a game between $N$ players defined as follows. For each player $\nu$, denote its action by $x_\nu\in\mathbb{R}^{n_{\nu}}$ and its payoff function by $\theta_\nu:\mathbb{R}^n\rightarrow\mathbb{R}$ where $n = \sum_{\nu=1}^{N}n_{\nu}$. Assume that each player $\nu$ has a strategy set $X_\nu(x_{-\nu}) \subseteq \mathbb{R}^{n_{\nu}}$ that could depend on the actions $x_{-\nu}$ of all other players. Hence, given the actions $x_{-\nu}$ of all other players, each player $\nu$ chooses a strategy $x_{\nu}$ that solves:
$$
\max_{x_\nu\in X_\nu(x_{-\nu})}\quad \theta_\nu(x_{\nu},x_{-\nu}).
$$
Suppose that:
\begin{enumerate}
\item There exists $N$ nonempty, convex and compact sets $K_\nu \subset \mathbb{R}^{n_\nu}$ such that for every $x \in \mathbb{R}^n$ with $x_\nu \in K_\nu$ for every $\nu$, $X_\nu(x_{-\nu})$ is nonempty, closed and convex, $X_\nu(x_{-\nu}) \subseteq K_\nu$, and $X_\nu$, as a point-to-set map, is both upper and lower semicontinuous.
\item For every player $\nu$, the function $\theta_\nu(\cdot, x_{-\nu})$ is quasi-concave on $X_\nu(x_{-\nu})$.
\end{enumerate}
Then a GNE exists.
\end{theoremNoNum}

\begin{IEEEproof}[Proof of Theorem \ref{thm:exist}]
We divide the proof into three cases, depending on the form of the market maker objective $\pi^M$.

\emph{{Case 1: $\pi^M = W_{soc}$.}} Here, we prove that a GNE always exists. Condition 1 in Debreu's Theorem requires strategy sets to be compact. It can be shown that the shift-factor matrix $H$ has rank $n-1$ for any power network and $\bone^\top$ is linearly independent from the rows of $H$. It then follows that the feasible region of injection is compact and hence the strategy set $S^M(q)$ of the market maker is also compact. Now, we turn our attention to the strategy sets of generators $S^G_k$. Though $S^G_k$ of generators are not compact, they can be restricted to some compact subset $[0, \bar{s}]$ since any equilibrium production $q_k^*$ can be upper bounded by some $\bar{s}$. To see the latter, first observe that, if $r_k^*$ is an equilibrium re-balancing quantity, then it is bounded from above since:
\begin{align*}
\int_0^{q_k + r_k^*} p_k (w_k) dw_k
=
a_k ( q_k + r_k^* ) - \frac{b_k}{2} ( q_k + r_k^* )^2,
\end{align*}
and that for large $r_k^*$, the quadratic term (which has a negative coefficient) dominates the linear term. Hence, suppose $r_k^* \leq \bar{r}$ for all $k$. Let $\bar{s} = a_k/b_k + (n-1)\bar{r}/b_k$. Note that, if $q_k^* > \bar{s}$, then the equilibrium price at node $k$ is:
\begin{align*}
p_k^*
%&= a_k - b_k (q_k^* + r_k^*)
%\\
&= a_k - b_k \left(q_k^* - \sum_{k'\neq k} r_{k'}^*\right)
%\leq a_k - b_k \left(q_k^* - (n-1)\bar{r} \right)
< 0.
\end{align*}
This is a contradiction since generator $k$ cannot be facing a negative price $p_k^* < 0$ and yet producing a positive quantity $q_k^* > \bar{s}$. For the rest of this proof, we shall assume that $S_k^G = [0, \bar{s}]$.

It is straightforward to show that our game satisfies conditions 1 and 2 in Debreu's Theorem. Condition 2 holds trivially since the generator and market maker payoffs are strictly concave over their respectively strategy sets. To see that condition 1 holds, choose $K_{\nu}$ in Debreu's Theorem in the following manner: (a) for each generator $k$, choose $K_{\nu} = S_k^G$; and (b) for the market maker, choose $K_{\nu} = \{r\in\mathbb{R}^n \;|\; |Hr| \leq f,\; \bone^\top r = 0\}$. It is clear that $K_{\nu}$ are nonempty, convex, and compact.

While the generator strategy sets $S_k^G$ are constant correspondences, the market maker strategy set $S^M(q)$ is a polytope that is linearly parametric. Thus, the strategy sets are both upper and lower semicontinuous in terms of players' actions.

\emph{{Case 2: $\pi^M = W_{res}$.}} Here, we prove that a GNE always exists. Observe that any equilibrium re-balancing quantity $r_k^*$ is bounded from above since:
\begin{align*}
\int_0^{q_k + r_k} p_k (w_k) dw_k - q_k p_k ( q_k + r_k )
=
a_k r_k - \frac{b_k}{2} ( r_k^2 - q_k^2 ).
\end{align*}
The rest of the proof is similar to that for case 1.

\emph{{Case 3: $\pi^M = W_{con}$.}} Here, we construct an example where GNE does not exist using the 2-node network in Figure~\ref{fig:2Node}. Our construction is based on the following lemma, proven in the Appendix.
\begin{lemma}
\label{lemma:conSurCounter}
Consider the 2-node network in Figure~\ref{fig:2Node}. Let $\pi^M = W_{con}$. Suppose $a_1 = a_2 = a $, $1 < b_1/ b_2 \leq 3$, $c_1 = c_2 = c$, and $f_{12}$ satisfies:
\begin{align}
\label{eq:conSurCounter}
\frac{a}{3 b_1 + 2 c} \ < \ f_{12} \ < \ \min \left\{ \frac{a}{b_2 + 2c}, \frac{a}{b_1}, f_0 \right\},
\end{align}
where:
$$ f_0 := \frac{a b_2 \Big[ b_1 + b_2 + c(3 -b_1/b_2)\Big]}{b_1 b_2 (b_1 + b_2) + b_1 (b_1 + 5 b_2) c + 2 (b_1 + b_2) c^2 }.$$
Then there does not exist a GNE.
\end{lemma}
The following parameter values: $a = 10$, $b_1 = 1.2$, $b_2 = 1$, $c =  1$ and $f_{12} = 2$, satisfy the conditions in the lemma and provides an example in which GNE does not exist.
%\begin{align*}
%f & < \frac{a}{b_2 + 2c} = 3.33, \\
%f & > \frac{a}{3 b_1 + 2 c} = 1.79, \\
%f & < \frac{a}{b_1} = 8.33, \\
%f & < f_0 = 2.76.
%\end{align*}
%$f = 2$ suffices!
\end{IEEEproof}

%%%%%%%%%%%%%%%%%%%%%%%%% 

%%%%%%%%%%%%%%%%%%%

\section{Regulatory objectives and market outcomes}
\label{sec:regObj}

Given the existence results in the previous section, we now move to analyzing the impact of regulatory objectives on the market outcomes.  To provide clear insight, we focus our analysis on the case of a the  2-node network in Figure~\ref{fig:2Node}, which represents a caricature of the situation in California. Though simple, this 2-node network is already enough to highlight significant differences in the impact of regulatory objectives.

We begin with a case of unbounded line capacities.  This allows us to consider a situation where the market equilibrium always exists for each regulatory objective.  Additionally, it highlights that the behaviors of the three regulatory objectives we are studying can differ dramatically even in the simplest of settings.  A proof of the following result is given in the appendix.

\begin{theorem}
\label{thm:regObj}
 Consider the 2-node network in Figure \ref{fig:2Node}.  Let $a_1 = a_2 := a$, $1 < b_1/b_2 \leq 3$, $c_1 = c_2 := c$ and $f_{12} = \infty$. Then a GNE exists for all $\pi^M \in \{ W_{soc}, W_{res}, W_{con} \}$. Moreover, the equilibrium re-balancing quantity $r_1^* = -r_2^*$ under the three regulatory objectives are as follows:
\begin{enumerate}[(a)]
\item If $\pi^M = W_{soc}$, then $r_1^* < 0$,
\item if $\pi^M = W_{res}$, then $r_1^* = 0$,
\item If $\pi^M = W_{con}$, then $r_1^* > 0$.
\end{enumerate}
\end{theorem}

Note that, even though there are no line constraints (i.e., $f_{12} = \infty$), the 2-node network is not equivalent to an aggregated market since the price at each node is a function of the local demand function at that node.

%--------Bose Edit---------------
\begin{figure*}[t]
\centering
\vspace{-.2in}
\subfloat[Social welfare]{\includegraphics[width=0.32\textwidth]{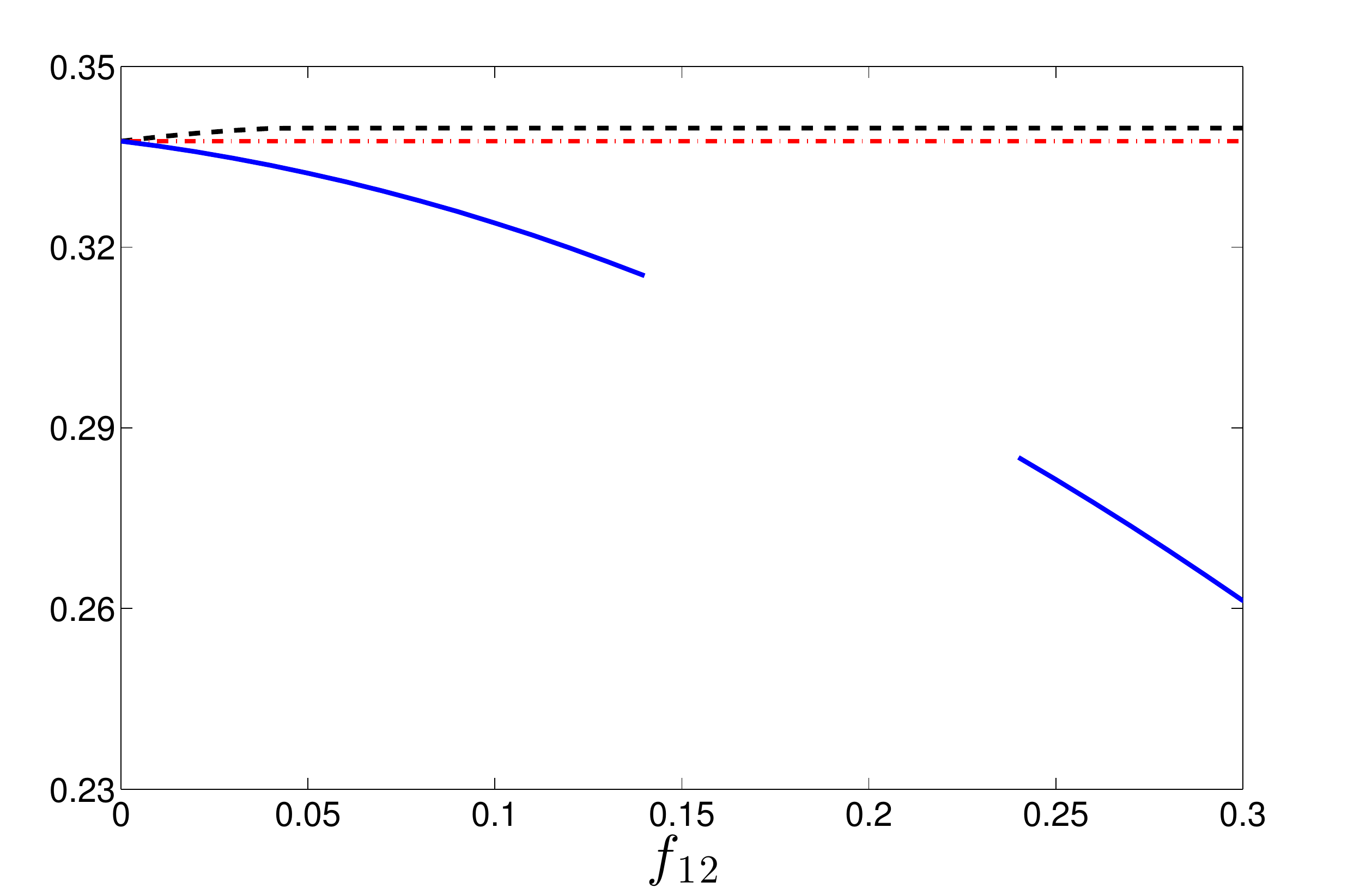}\label{fig:plotSocWel}}
\subfloat[Consumer surplus]{\includegraphics[width=0.32\textwidth]{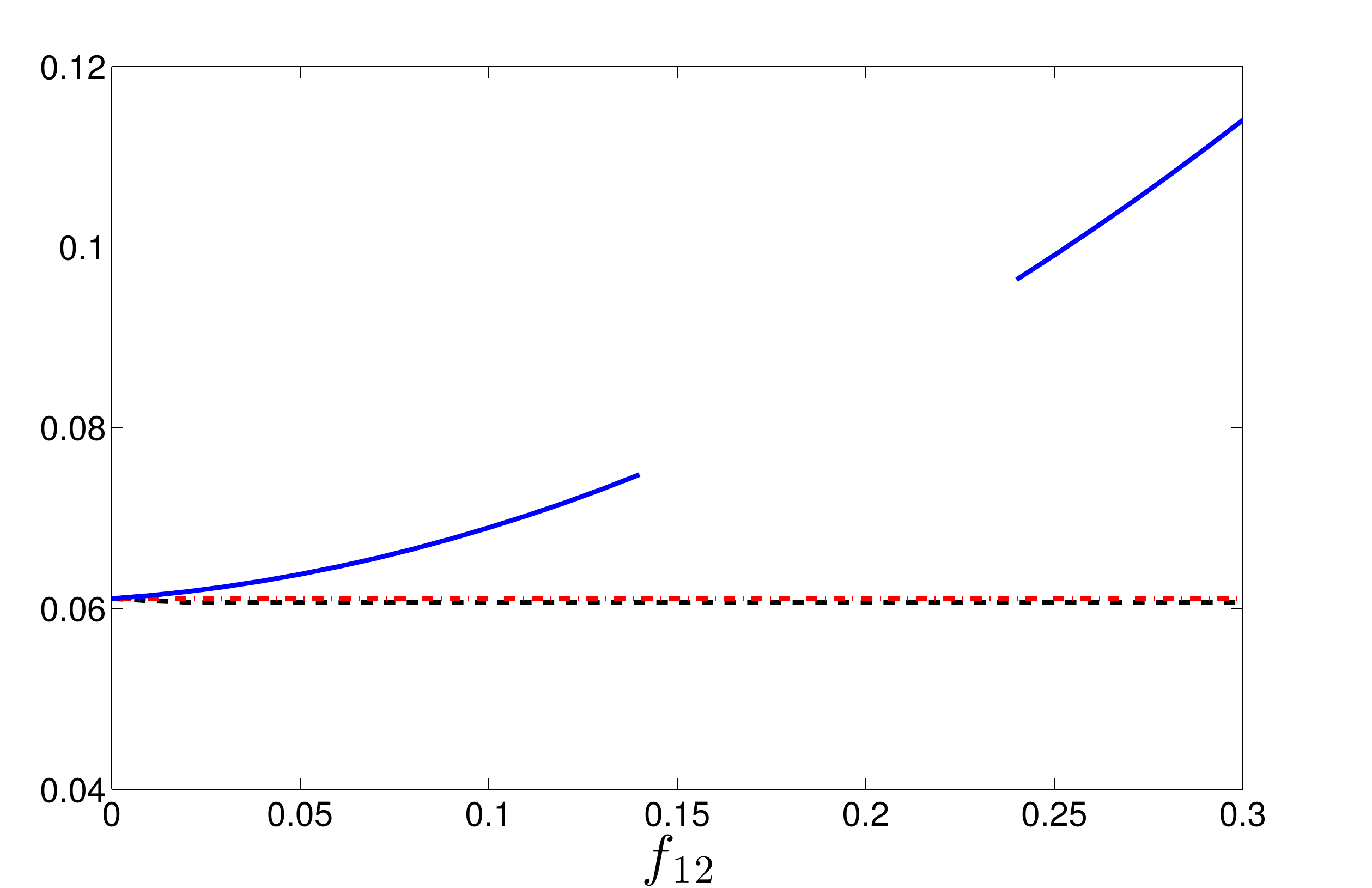}\label{fig:plotConSur}}
\subfloat[Generator profits]{\includegraphics[width=0.32\textwidth]{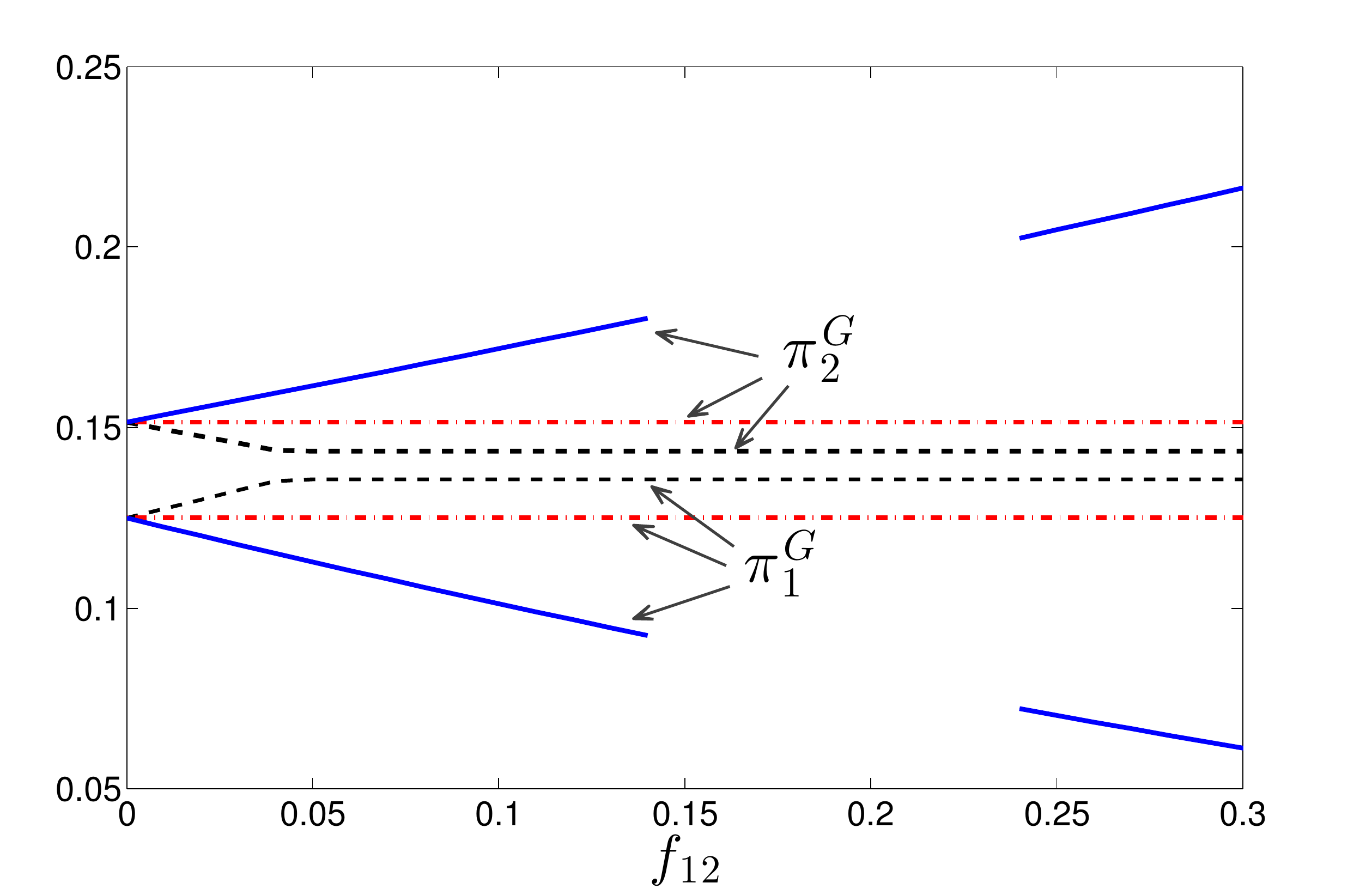}\label{fig:plotProfits}} \\
\subfloat{\includegraphics[width=0.45\textwidth]{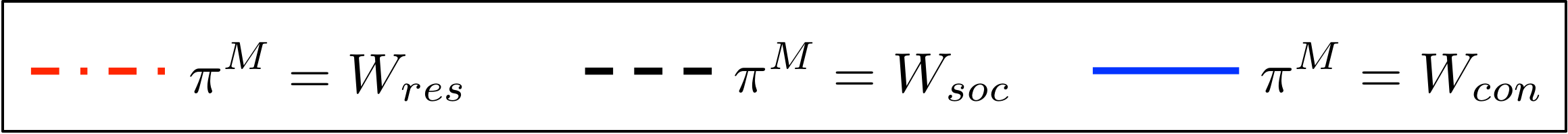}}
\caption{Plots of various quantities at equilibrium with varying line capacities $f_{12}$ in the 2-node network in Figure \ref{fig:2Node}. Parameters chosen for this example are: $a_1 = a_2 = 1$, $b_1 = 1$, $b_2 = 0.65$, $c_1 = c_2 = 1$.}
\end{figure*}
%----------Bose Edit finish---------

Our result illustrates how a simple market can exhibit very different equilibria under different regulatory objectives. In particular, though all three market maker objectives are motivated qualitatively by the identical goal of maximizing consumer benefit; one results in flow going north, one in flow going south, and one in no flow between the nodes. So, the exact choice of how costs are reflected in the objective is a significant determinant of how generators behave in the market, which affects the equilibrium power flows and system efficiency dramatically. Hence the market design question is important in the operation of a deregulated market. Although Theorem~\ref{thm:regObj} assumes that the line capacity $f_{12} = \infty$, our numerical calculations indicate that the sign of $r_1^*$ exhibit the same properties even under a binding line constraint.

To further emphasize the significance of the market maker objective on the efficiency of the market, we compare the social welfare (\figurename~\ref{fig:plotSocWel}), consumer surplus (\figurename~\ref{fig:plotConSur}), and generator profits (\figurename~\ref{fig:plotProfits}), at the unique equilibrium under each of the three market maker objectives as the line capacity $f_{12}$ is increased. Here, we choose the parameters in the following manner: $a_1 = a_2 = 1$, $b_1 = 1$, $b_2 = 0.65$, and $c_1 = c_2 = 1$; but the qualitative features in the plots continue to hold for other parameter values that we experimented with. For the case where $\pi^M = W_{con}$, the gap in the plot indicates that equilibrium does not exist for those values of $f_{12}$. The plots reveal the counter-intuitive phenomena that: increasing the line capacity could decrease social welfare if $\pi^M = W_{con}$. There is also a clear tradeoff between market maker objectives: having $\pi^M = W_{soc}$ leads to higher social welfare but lower consumer surplus versus having $\pi^M = W_{con}$.

The three market maker objectives also lead to completely different scaling of generator profits as the line capacity $f_{12}$ is increased -- generator $G_1$ benefits from line expansion when $\pi^M = W_{soc}$ but generator $G_2$ benefits from line expansion when $\pi^M = W_{con}$. This implies that, although the market maker objective is only used in the short-term market, it also has implications on long-term incentives to expand transmission.

%\begin{figure}
%\centering
%\subfloat[Equilibrium flow $r_1^*$.]{\includegraphics[width=0.23\textwidth]{Plots/optimalFlow.eps}\label{fig:optimalFlow}}
%\subfloat[Equilibrium social welfare.]{\includegraphics[width=0.23\textwidth]{Plots/socialWelfare.eps}\label{fig:socialWelfare}}
%\\
%\subfloat[Equilibrium output $q_1^*$.]{\includegraphics[width=0.23\textwidth]{Plots/outputQ1.eps}\label{fig:outputQ1}}
%\subfloat[Equilibrium output $q_2^*$.]{\includegraphics[width=0.23\textwidth]{Plots/outputQ2.eps}\label{fig:outputQ2}}
%\\
%\subfloat[Profit of generator $G_1$.]{\includegraphics[width=0.23\textwidth]{Plots/profitG1.eps}\label{fig:profitG1}}
%\subfloat[Profit of generator $G_2$.]{\includegraphics[width=0.23\textwidth]{Plots/profitG2.eps}\label{fig:profitG2}}
%\caption{Equilibrium values for varying line capacities $f$ in the 2-node example. Here, we set $a_1 = a_2 = 1$, $b_1 = 1$, $b_2 = 0.65$, $c = 1$.}
%\end{figure}

%\input{residualSocialSurplus.tex}
%\input{consumerSurplus.tex}

\section{Conclusion and future work}

In this work we introduce a new networked Cournot model for studying the impact of regulatory objectives on the outcomes in electricity markets. In particular, the model we introduce formulates a game between the electricity market maker (or market operator) and generators.  Within this game, our main results explore the contrasts between three natural market maker objectives -- social welfare, residual social welfare, and consumer surplus. The results in this paper reveal that the design of the market has significant implications on both the existence and form of equilibria. In particular, equilibria might not exist when the market maker maximizes the consumer surplus and the network is capacity constrained. Further, even when equilibria exists, the equilibrium allocation of power flows can be completely different under the three market maker objectives. Hence, the results in this paper highlight that design of market maker objective is delicate and needs to be further investigated in a principled manner.

%\adam{where do you want to discuss the impact of the results for the more general network Cournot literature?}

 \appendix
Here we present the proofs of Lemma \ref{lemma:conSurCounter} and Theorem \ref{thm:regObj}. These results are specific to the 2-node network in Figure \ref{fig:2Node}. Hence, we simplify the notation by defining $r := r_1 = -r_2$. Furthermore, we drop the subscript in $f_{12} := f$.\footnote{The notations $f$ and $r$ in this proof should not be confused with the vectors in Section \ref{sec:formulation}.}

By applying the assumption that $a_1 = a_2 := a$, we can write the derivatives of the generator profits with respect to their production quantities as:
\begin{align}
	\frac{\dd\pi_1}{\dd q_1} & \ = \ (a - b_1 r) - 2(b_1 + c) q_1 \label{eq:dq1},\\
	\frac{\dd\pi_2}{\dd q_2} & \ = \ (a + b_2 r) - 2(b_2 + c) q_2 \label{eq:dq2}.
\end{align}
We make repeated references to these expressions throughout the proofs.

\subsection*{Proof of Lemma \ref{lemma:conSurCounter}:}
Here $\pi^M(q_1, q_2, (r, -r)) = W_{con}(q_1, q_2, (r, -r))$. From equation~\eqref{eq:con.G}, we get:
$$\pi^M((q_1, q_2), (r, -r)) = \underbrace{\frac{b_1}{2} (q_1 + r)^2 +  \frac{b_2}{2} (q_2 - r)^2}_{:=\Pi(r)/2}.$$
The market maker maximizes $\Pi(r)$ subject to $-q_1 \leq r \leq q_2$ and $-f \leq r \leq f$. Our proof technique is to completely characterize all possible equilibria $(q_1^*, q_2^*, r^*)$ and the conditions on $f$ that lead to each of the equilibria. Since those conditions on $f$ do not contain the relation in~\eqref{eq:conSurCounter}, we then infer that GNE does not exist when $f$ satisfies~\eqref{eq:conSurCounter}.

We divide our analysis into two cases based on whether $f \geq a/(b_2 + 2c)$ or $f < a/(b_2 + 2c)$.  The first case can be interpreted as the scenario in which network constraints are not tight.

\textbf{Case 1:  $f \geq a/(b_2 + 2c)$}. Here, we show that a GNE always exists by constructing one. In particular, we construct a GNE such that $r^* = q_2^*$. Note that, since $\Pi$ is convex, its maximizers occur at $-q_1^*$, $q_2^*$, $-f$, or $f$. By using $b_1 > b_2$, we can check that, for any $q_1^*, q_2^* \geq 0$, we have:
$$\Pi(q_2^*) \geq \Pi( \max \{-q_1^*, -f \} ).$$
Since $a + b_2 r^* = a + b_2 q_2^* \geq 0$, we can solve for $q_2^*$ by setting $\frac{\dd\pi_2}{\dd q_2} \Big\vert_{q_2^*} = 0$ in~\eqref{eq:dq2}, which gives:
$$ q_2^* = r_* =  \frac{a}{b_2 + 2c}.$$
Now note that $q_2^* < f$ which verifies that $q_2^*$ maximizes $\Pi(r)$ over $r \in[-q_1^*,q_2^*]\cap[-f,f]$. Next, using~\eqref{eq:dq1} to solve for $q_1^*$ gives:
$$ q_1^* =
\begin{cases}
a \frac{2c + b_2 - b_1}{2(b_1 + c)(b_2 + 2c)}, & \textit{if}\; b_1 < b_2 + 2c,
\\
0, & \textit{otherwise}.
\end{cases}
$$
This defines a GNE.

\textbf{Case 2: $f < a/(b_2 + 2c)$}. First, we argue that any equilibrium must satisfy $q_2^* \geq f$. Suppose there exists an equilibrium with $q_2^* < f$. The analysis in case 1 implies that $r^* = q_2^*$. However, the first-order condition for generator 2 (c.f.~\eqref{eq:dq2}) implies that $q_2^* = r^* = a/(b_2 + 2c) > f$ which is a contradiction. Hence, any equilibrium must satisfy $q_2^* \geq f$.

Recall that $\Pi$ is strictly convex. The condition that $q_2^* \geq f$ imply that the maximizers of $\Pi$ can only occur at $-q_1^*$, $-f$, or $f$. We consider each case separately. Due to lack of space, we only give the proof of the case where $-f \leq -q_1^*$ and $r^* = +f$ in this paper. However, the approach for the other cases is similar.

%\begin{enumerate}
%\begin{figure*}[!h]
%\centering
%{\includegraphics[width=0.4\textwidth]{cond1.png}}
%\end{figure*}
%\emph{When $-f \leq - q_1^*$ and $r^* = +f$:}
Suppose $-f \leq - q_1^*$ and $r^* = +f$. From \eqref{eq:dq1} and \eqref{eq:dq2}, we have:
\begin{align*}
q_2^* = \frac{a+b_2 f}{2(b_2 + c)}, \quad q_1^*  = \begin{cases} \frac{a-b_1 f}{2(b_1 + c)}, & \textit{if}\; f \leq \frac{a}{b_1},\\
0, & \textit{otherwise}.
\end{cases}
\end{align*}
For this case, we need the following conditions to be satisfied: (a) $q_2^* \geq f$, (b) $q_1^* \leq f$, and (c) $\Pi(+f) \geq \Pi(-q_1^*)$. We derive conditions on $f$ for (a), (b) and (c) to hold. It can be checked that $f < a/(b_2 + 2c)$ implies (a) is always satisfied. To deal with conditions (b) and (c), we consider the two possibilities separately: (i) $f \leq {a}/{b_1}$, and (ii) $f > a/b_1$.

(i) Suppose $f \leq {a}/{b_1}$. Then (b) $q_1^* \leq f$ if and only if:
$$ f \geq \frac{a}{3 b_1 + 2 c}.$$
Also, (c) $\Pi(+f) \geq \Pi(-q_1^*)$ is true if and only if the following quantity is non-negative.
\begin{align*}
& \Pi(+f) - \Pi(-q_1^*)\\
& \quad = b_1 (q_1^* + f)^2 + b_2 (q_2^* - f)^2 - b_2(q_1^* + q_2^*)^2 \\
%& \quad = b_1 (q_1^* + f)^2 + b_2 \Big[ (q_2^* - f)^2 - (q_1^* + q_2^*)^2 \Big] \\
%& \quad  = b_1 (q_1^* + f)^2 + b_2 \Big[ (q_2^* - f + q_1^* + q_2^*) (q_2^* - f - q_1^* - q_2^*) \Big] \\
& \quad = \underbrace{(q_1^* + f)}_{\geq 0}\underbrace{\Big[ b_1(q_1^* + f) - b_2 (2 q_2^* - f + q_1^*) \Big].}_{:=\lambda}
\end{align*}
Substituting the expressions for $q_1^*$ and $q_2^*$ for this case, it can be verified that $\lambda \geq 0$ if and only if:
$$ f \geq \frac{a b_2 \Big[ b_1 + b_2 + c(3 -b_1/b_2)\Big]}{b_1 b_2 (b_1 + b_2) + b_1 (b_1 + 5 b_2) c + 2 (b_1 + b_2) c^2 } := f_0.$$

(ii) Suppose $f > {a}/{b_1}$. Then (b) $q_1^* = 0 \leq f$ is trivially satisfied.
Also, (b) $\Pi(+f) \geq \Pi(-q_1^*)$ if and only if $\lambda \geq 0$, where:
\begin{align*}
\lambda
& = b_1(q_1^* + f) - b_2 (2 q_2^* - f + q_1^*)\\
%& = b_1 f - b_2 (2 q_2^* - f) \\
%& = (b_1 + b_2) f - 2 b_2 q_2^*\\
& = (b_1 + b_2) f - 2 b_2 \Big[\frac{a+b_2 f}{2(b_2 + c)}\Big] \\
%& = \Big(b_1 + b_2 - \frac{b_2^2}{(b_2 + c)}\Big) f - \frac{a b_2 }{(b_2 + c)} \\
& = \frac{b_1 b_2 + c(b_1 + b_2)}{(b_2 + c)} f - \frac{a b_2 }{(b_2 + c)}.
\end{align*}
Now, we also have:
$$ f \geq \frac{a}{b_1} > \frac{a}{b_1 + c (1 + b_1/ b_2)} \implies \lambda \geq 0.$$

By working through the other cases in a similar manner, we discover that there exists a GNE if and only if:
\begin{enumerate}
\item $f \geq a/(b_2 + 2c)$; or,
\item $f < a/(b_2 + 2c)$, $f \leq a/b_1$, $f \geq a/(3b_1 + 2c)$, and $f \geq f_0$; or,
\item $f < a/(b_2 + 2c)$ and $f > a/b_1$; or,
\item $f < a/(b_2 + 2c)$, $f \leq a/b_1$, $f \leq a/(3b_1 + 2c)$, and $f \leq f_1$,
\end{enumerate}
where $f_1 := \frac{a c (b_1 - b_2)}{b_1 b_2 (b_1 + b_2) + c (b_1^2 + b_2^2)}$. Since the relation in \eqref{eq:conSurCounter} is not contained in any of the above cases, this completes the proof of Lemma \ref{lemma:conSurCounter}. $\hfill\blacksquare$

%\subsection{An example where eq. does not exist}
%To construct such an example, first let's have the following set of inequalities:
%\begin{align*}
%f & < \frac{a}{b_2 + 2c}, \\
%f & > \frac{a}{3 b_1 + 2 c}, \\
%f & < \frac{a}{b_1}, \\
%f & < f_0.
%\end{align*}
%
%The first one forces one among conditions (1) and (3). The second one avoids condition (3) and leaves with the choice of condition (1). Then third one leads to option (I). Now, making $f < f_0$, we violate even that condition of equilibrium. In particular, if such an $f$ exists, then there is NO possible equilibrium. Consider the case $a = 10$, $b_1 = 1.2$, $b_2 = 1.0$, $c =  1$. In this case, we have
%\begin{align*}
%f & < \frac{a}{b_2 + 2c} = 3.33, \\
%f & > \frac{a}{3 b_1 + 2 c} = 1.79, \\
%f & < \frac{a}{b_1} = 8.33, \\
%f & < f_0 = 2.76.
%\end{align*}
%$f = 2$ suffices!

%%%%%%%%%%%%%%%%%%%%%%%%%%
%%%%%%%%%%%%%%%%%%%%%%%%%%
%%%%%%%%%%%%%%%%%%%%%%%%%%

\subsection*{Proof of Theorem \ref{thm:regObj}:}

\noindent\emph{Case (a): $\pi^M = W_{soc}$.}
Simplifying the expression for $W_{soc}$ in~\eqref{eq:soc.G} gives:
\begin{align*}
2 \pi^M(q_1, q_2, (r,-r))
& = \underbrace{-b_1 (q_1 + r)^2 - b_2 (q_2 -r)^2}_{:=\Pi(r)}\\
& \qquad + 2 a (q_1 + q_2) - 2 c_1 q_1^2 - 2 c_2 q_2^2.
\end{align*}
Maximizing $\pi^M(q_1, q_2, r)$ is equivalent to maximizing $\Pi(r)$ over $r \in [-q_1, +q_2]$. It can be checked that $\Pi(r)$ is always maximized at an interior point and hence, at equilibrium, the quantities $q_1^*, q_2^*, r^*$ satisfy:
\begin{align}
\label{eq:rStar}
r^* = \frac{b_2 q_2^* - b_1 q_1^*}{b_1 + b_2}.
\end{align}
To compute $q_1^*$ and $q_2^*$, note that there are four possible configurations of equilibria depending on the signs of $a-b_1 r^*$ and $a+ b_2 r^*$. We deal with each case separately.
\begin{enumerate}[(i)]

\item $a-b_1 r^* < 0$, $a + b_2 r^* < 0$: From \eqref{eq:dq1} and \eqref{eq:dq2}, it follows that $q_1^* = q_2^* = r^* = 0$. But then we have $a - b_1 r^* = a > 0$ and hence a contradiction. Hence, an equilibrium of this form does not exist.

\item $a-b_1 r^* < 0$, $a + b_2 r^* \geq 0$: From \eqref{eq:dq1} and \eqref{eq:dq2}, we have $q_1^* = 0$, and $q_2^* = (a + b_2 r^*)/(2b_2 + 2c)$. Substituting this into \eqref{eq:rStar} and simplifying, we get:
$$ r^* = \frac{b_2}{b_1+b_2} q_2^* = \frac{a b_2}{2(b_1+b_2)(b_2 + c) - b_2^2}.$$
But it can be checked that $a - b_1 r^* \geq 0$ which is a contradiction. Hence, an equilibrium of this form does not exist.

\item $a-b_1 r^* \geq 0$, $a + b_2 r^* < 0$: From \eqref{eq:dq1}, \eqref{eq:dq2} and using arguments similar to the last case, we have $q_2^* = 0$ and:
$$ r^* = \frac{- a b_1}{2(b_1+b_2)(b_1 + c) - b_1^2}.$$
Again, it can be checked that $a + b_2 r^* \geq 0$ which is a contradiction. Hence, an equilibrium of this form does not exist.

\item $a-b_1 r^* \geq 0$, $a + b_2 r^* \geq 0$: For this case the triplet $(q_1^*, q_2^*, r^*)$ satisfies the relation in \eqref{eq:rStar} and:
\begin{align*}
q_1^* = \frac{a - b_1 r^*}{2(b_1 + c)}, \ \text{and} \ q_2^* = \frac{a + b_2 r^*}{2(b_2 + c)}.
\end{align*}
Solving these linear equations, we obtain:
$$r^* = \frac{a c (b_2 - b_1)}{  (b_1 + b_2) (b_1 b_2 + 2 c^2) +  c (b_1^2 + b_2^2 + 4 b_1 b_2 )} < 0.$$
With some algebraic manipulations, it can be shown that indeed $a-b_1 r^* \geq 0$ and $a + b_2 r^* \geq 0$. This defines an equilibrium.
\end{enumerate}
This proves the claim in Theorem \ref{thm:regObj}(a).

\noindent\emph{Case (b): $\pi^M = W_{res}$.}
Simplifying the expression for $W_{res}$ in~\eqref{eq:res.G} gives:
$$2 \pi^M (q_1, q_2, (r, -r)) = - (b_1+b_2) r^2 + b_1 q_1^2 + b_2 q_2^2.$$
Since $\pi^M$ is strictly concave in $r$, it is maximized at $r^* = 0$. The resulting equilibria values for $q_1^*$ and $q_2^*$ can be computed from the generator profits. This proves the claim in Theorem \ref{thm:regObj}(b).

\noindent\emph{Case (c): $\pi^M = W_{con}$.}
Since $f > a/(b_2 + 2c)$, this corresponds to case 1 in the proof of Lemma~\ref{lemma:conSurCounter}. Hence, equilibrium always exists and we have:
$$r^* =  \frac{a}{b_2 + 2c} > 0.$$
This proves the claim in Theorem \ref{thm:regObj}(c), which completes the proof of the theorem. $\hfill\blacksquare$

%%%%%%%%%%%%%%%%%%%%%%%%%%%%%%%%%%%%%%%%

\singlespacing
\bibliography{PowerMarket}
\end{document}